# PROFIT WARNINGS AND STOCK RETURNS: EVIDENCE FROM MOROCCAN STOCK EXCHANGE

**Ilyas EL GHORDAF (PhD)**

University Mohammed First, Oujda, Morocco

**Email:** elghordaf.ilyass@gmail.com

**Abdelbari EL KHAMLICHI (PhD)**

LERSEM, ENCG

Chouaib Doukkali University, El Jadida, Morocco

**Email:** abdelbari.el.khamlichi@gmail.com

**Abstract**

There is an important literature focused on profit warnings and its impact on stock returns. We provide evidence from Moroccan stock market which aims to become an African financial hub. Despite this practical improvement, academic researches that focused on this market are scarce and our study is a first investigation in this context. Using the event study methodology and a sample of companies listed in Casablanca Stock Exchange for the period of 2009 to 2016, we examined whether the effect of qualitative warning is more negative compared to quantitative warnings in a short event window. Our empirical findings show that the average abnormal return on the date of announcement is negative and statistically significant. The magnitude of this negative abnormal return is greater for qualitative warnings than quantitative ones.

# I-INTRODUCTION

Financial communication is one of the first obligations of listed companies in any stock exchange market. In order to relay financial information to all interested parties, companies use many ways such as annual reports, or earnings conference calls, etc.. However, public companies must adopt a voluntary financial communication strategy which exceeds legal requirements in terms of information. It must take into account the economic and strategic dimension of the firms within the limits permitted by the conditions of competition and respect for business secrecy. In this regard, the Casablanca Stock Exchange has witnessed an increase in profit warning in the recent past years with 23 warnings in year 2015. Profit warnings are meant to reduce the effect of surprise at the moment of the real earnings announcements when the firm's forthcoming earnings will not meet the previous expectations earnings.

This event is a visible message to financial market declaring an important negative change in the performance of a company. Consequently, if the market is efficient, asset prices should respond rapidly to the new information (Brealy and Myers 2000).

As a tool of financial communication, profit warnings improve transparency between listed companies and investors in the Stock Exchange. However, managers can make the choice between qualitative and quantitative announcements. The quantitative warnings give the exact estimation of earnings or intervals, while the qualitative ones indicate only that profit will fall under current expectations without any specific quantification of the new earnings and is considered to be a less credible signal (Heesters 2011).These two types of profit warnings offer the possibility to test not only if the market responds to news, but also if the extent of the reaction is related to the quality of the information released.

Several studies have investigated the market reaction to profit warnings. In a recent paper Dayanandan et al. (2017) found that profit warnings reduce information asymmetry and increase trading volumes, even though stock market reactions are more adverse for bad news. Actually, there is considerable literature on profit warnings and its impact on financial outcomes (Aubert and Louhichi 2015). Event study methodology was also applied for some African stock markets, especially South Africa and Nigeria (Olowe 2011, Rabin 2015). But the studies are scarce due to the limited data availability for African markets as reported by Ellis and Keys (2014).

To the best of our knowledge, none of the studies focused on the Moroccan stock exchange, which is considered as the third largest stock market after South African and Nigerian stock exchanges. With some initiatives (such as Casablanca Finance City, launched in 2010), Morocco aims to become an African financial hub.

The aim of this article is to analyze the market reaction to profit warnings issued by companies listed at Casablanca Stock Exchange for the period of 2009 to 2016. Particularly this work examines the difference in impact between qualitative and quantitative warnings.

The paper is structured as follows. We start by presenting the literature review followed by testable hypotheses. Then, we describe our data and event study methodology. In section four, the findings of the article are presented and discussed. Finally, the fifth section presents our conclusions.

# II- LITTERATURE REVIEW AND HYPOTHESES DEVELOPPEMENT

## II-1 WHAT IS A PROFIT WARNING

When a firm forecasts that its earnings will not achieve analyst expectations, the profit warning is made to adjust available expected results before the public release. If no profit warning is released, the impact of this bad news can be very huge at the date of earnings announcements due to the effect of surprise.

By disclosing this information, companies share the information with participants in the stock exchange that they will not reach the market expectations regarding the performance.

On the other hand, these warnings are typically made around the end of financial period, but prior half-annual or annual earnings report (Elayan and Pukthuanthong, 2009). Consequently, not to warn the real financial situation of the company lowers its valuation and liquidity, which increases the cost of capital.

To avoid a large decline in the stock market, the profit warning in the Moroccan Stock Exchange must obey the following criteria:
- **Timing:** A profit warning reaches its objective when published in the available time. Announcing a profit warning a few days before legal publication can't mitigate the sudden impact that may suffer the stock price. Furthermore, the publication of this information without any delay reduces the risk of insider trading. The Moroccan Financial Market Authority has established rigorous rules concerning insider trading activity. Especially, these regulations require that all insider stock transactions must be reported as soon as possible.
- **Clarity:** The profit warning must explain the reasons of decrease in profit by bringing sufficient information about the environment of the company, and eventually the action taken or proposed to deal with the new situation.
- **Equal access to information:** An eventual drop in results is considered like important information that must be public for all the investors without any discrimination through official media.

### II-1-1 WHY FIRMS ISSUE PROFIT WARNINGS?

Several studies have discussed why firms issue profit warning. They identified many reasons for companies to announce earnings-related warnings: to instance to prevent from huge drop in stock price, to avoid shareholder lawsuit, to maintain a good reputation, features of regulation in the market, etc.

While all these motivations are plausible, information asymmetry and reputation cost appear to be the most common justifications related to the topic.

#### II-1-1-1 Information asymmetry

The management of companies can possess private information about the firm's operations and future aspects due to the separation of ownership from control (Elayan and Pukthuanthong, 2009). This information asymmetry is a current problem in corporate finance when managers have much better information about firms than investors. To reduce this imperfect knowledge, the profit warning is announced by the directors as a message to public that future earnings will not meet market forecasts. It informs investors and shareholders about decline in expected results (Francoeur et al. 2008). Like other voluntary disclosures, profit warnings are meant to reduce information asymmetry, and improve liquidity in the market (Dayanandan et al. 2017).

Furthermore, Helbock and Walker (2003) made two alternative explanations to announce profit warning in UK. The first justification is that the English investors need more voluntary publication such as profit warning because listed companies are not required to disclose quarterly financial information. Consequently, the frequent publication of legal financial statements reduces the information asymmetry and decrease the probability of issuing profit warning.

In the same time, these authors exhibit a second characteristic of British firms in terms of ownership structure where the role of institutional investors is higher than other countries. The pressure of these shareholders leads companies to be more transparent and encourage them to warn about any change in the expected results.

### II-1-1-2 Reputation cost

Reputation is a critical factor to attract the financial resources required to enhance the development of companies. In the case of profit warnings, managers want to maintain a good relationship with investors through transparent communication. Disclosure of the information should drive financial analysts to revise quickly their expected earnings especially for the firms that are less followed by the market (Holland and Stoner 1996). It's therefore necessary in the interest of credibility of information, provided by the companies, to notify all new information related to the future of their business. Moreover, transparent communication contributes in maintaining the reputation in the credit market and this can be a huge pressure to warn bad news.

On the other side, a firm possessing good news wants to inform the market with the aim to be different from its competitors (Skinner 1994) in terms of integrity and this factor can give important strengths to companies.

### II-1-2 CLASSIFICATION OF PROFIT WARNING

The announcement of profit warning can take two forms: either communicate to the market the new forthcoming earnings announcements (*quantitative warnings*) or simply warn investors that the objective will not be achieved without giving further details (*qualitative warnings*). The main difference between these two groups of warnings is that the quantitative one includes numerical earnings forecasts while the qualitative one contains only a downward tendency statement.

To shed light on what motivates this choice, we report to the following studies: The first study conducted by Soffer et al. (2000) found that both litigation risk and market reaction argument push managers to adopt a full disclosure strategy. They argue that partial disclosure of bad news can cause negative reaction of the market at the earning announcements date. Therefore, to avoid shareholder lawsuits, companies are required to release important information to investors as quickly as possible.

The second study initiated by Bulkley and Herrerias in 2005 assumes that psychological biases (for more details see Shefrin 2001) leading to excessive optimism and overconfidence are enhanced when the precision of signal decreases. In the case of announcing profit warnings, they predict a high level of market underreaction for qualitative statement since it is a less precise information compared to a quantitative warning. The underreaction continues for three months for quantitative warnings with significant abnormal returns of -1,98%. Whereas, the market underreaction continues for approximately six months for qualitative announcements with significant abnormal returns of 11,78%.

Finally, in the Netherlands, Church and Donker (2010) suggest that external factors are responsible for the warning as general production report and firms hold internal factors responsible for the warning as specific earning forecasts. The findings highlight that for successive warning the reaction is less important for firms adopting a full disclosure policy.

### II-2 DEVELOPMENT OF TESTABLE HYPOTHESES

In order to verify if the Moroccan Stock Market reacts significantly or not to a profit warning announcement and regarding the role of the quality of the information released on the market reaction, two hypotheses will be presented and discussed.

### II-2-1 PROFIT WARNINGS AND STOCK RETURNS

Announcing a profit warning reveals that the earnings are below the market forecasts. This bad news can affect the firm's stock price because investors evaluate the company's outlook based on the analysis of the firm's financial statements and external environment. Therefore, the market reaction will be negative at the date of the publication of warning. The investors become worried about the power of the company to be competitive in the future after the release of the new estimation. Consequently, the firm's value decreases as results from the deterioration of investment opportunities in their securities.

Several studies have been conducted to test the effect of profit warning on stock returns around the event window (Jackson and Madura 2003; Bulkley and Herrerias 2005; Elayan and Pukthuantong 2009). They found significant negative returns at the date of a profit warning announcement. However, other researches in behavioral finance provide evidence that investors over react or under react to the warnings announcements (Easterwood and Nutt 1999; Hede 2012; Dons and Sletnes 2003). The major justification for this phenomenon is that markets continue to drift downward after a bad news in contrary to the efficient market hypothesis where stocks prices change faster to reflect new information (Fama, 1970).

Thus, our first hypothesis is the following:

  **H1**: *Publishing a profit warning in the Moroccan Stock Exchange Market leads to a negative reaction around the announcement date.*

### II-2-2 TYPE OF PROFIT WARNING AND STOCK RETURNS

The second hypothesis of this study is about information content in profit warning. We will examine the role of high level of disclosure on shareholder value. The qualitative warning is considered as worse news because they contain less information compared to the quantitative one. Therefore, the market rewards companies that release detailed warnings. Prior studies (e.g. Skinner, 1994; Clare, 2001; Helbok and Walker, 2003; Bulkley and Herrerias, 2004 and Collet, 2004) have found out that abnormal returns are lower for firms that provide only general statement. If differences are detected between these two types of announcements it is expected that stock price adjust quickly to equilibrium when numerical forecasts are mentioned, and thereby reducing the information asymmetry between shareholders and management.
  **H2**: *The type of profit warning impacts the extent of the reaction in the event window.*

### III- DATA AND METHODOLOGY

In this section, we will provide a description of data sets to analyze the reaction of stock returns to profit warnings announcements in Morocco including some descriptive statistics.

We will also describe the steps of the event study methodology which is used to quantify, firstly, the whole impact of the event (profit warning), and, secondly, the market reaction to quantitative and qualitative statements by splitting the sample into two subsamples. The objective of this division is to examine whether differences are detected between these two types of profit warning.

### III-1 DATA

To test the impact of profit warning on stock price in the Moroccan Stock Exchange, we need to build our own database and retrieve information concerning this topic from different sources mainly online ones.

### III-1-1 DATA COLLECTION

We collect profit warning announcements for the period of January 2009 to July 2016 from the website of the Moroccan Financial Market Authority (AMMC), which includes all press releases concerning profit warning issued by listed companies on the Casablanca Stock Exchange. The AMMC's rules require that profit warning should be issued when the firm expects that its actual result will be significantly low in comparison with the history of results of issuer or the forecast disclosed by the company.

For each press release, we retrieve information concerning the announcement date and the type of warning. The profit warning is considered as qualitative when company declares that current expectation deviates from the financial position without giving numerical guidance.

The daily share price and information concerning the sector of the sample are obtained from the website of the Casablanca Stock Exchange. The initial sample contains 65 profit warnings. However, each announcement must respect the following criteria to be included in the final sample.

- Repeated warnings issued within two years of a first announcement are eliminated from the sample to avoid overlapping multi-month returns (Jackson and Madura 2007);
- For each company, the daily share price used in the event study needs to be available in the database of the Casablanca Stock Exchange;
- Profit warnings announced in the same time of confounding news such as bond issue or possible takeover are excluded from the sample to quantify only the effect of this event on share prices.

### III-1-2 DESCRIPTIVE STATISTICS:

After the above method, the final sample contains 42 profit warnings accessible to analysis. The tables below describe the profit warning by years and sectors for total sample.

As shown by table 1 the number of profit warning is low at the beginning of study period. However, in 2016 the number of announcement achieved its summit containing 31% of the total sample. This finding can be justified by two factors. First, listed companies fear the sanctions of the market authority in case of non-compliance with legal rules concerning the communication of important information that may have incidence on their stock prices.

Secondly, the quality of financial reporting improved significantly over the last years to take into account the requirement of the Moroccan Capital Market and international practices.

Table 2 shows that 26% of profit warnings are issued by firms operating in two major sectors: Mining and computer services. This situation is explained by the fall in metal prices in the international mining market that may not cover the cost of the exploitation and consequently, a decrease in profitability in terms of earnings.

For materials, software &computers services sector, the number of profit warnings is high due to slowdown in domestic consumption of their products but also, the prices drop substantially in a very competitive technology market.

### III-2 METHODOLOGY

The event study is used to investigate the effect of a profit warning announcement on the value of firms. This methodology seeks to determine whether there is a significant reaction associated with profit warning. According to the efficient market hypothesis, the impact of such event will be reflected immediately in the stock prices of the company. Consequently, we can observe the economic effect over a period of time (event window).

Three steps are necessary to conduct an event study: Identify the event, estimate parameters from the model market to measure normal and abnormal returns. Finally, test the significance of abnormal returns around the event date.

### III-2-1 EVENT DEFINITION

To conduct an event study we need to convert calendar time to event time. This time-line (Skerpnek and Lawson 2001) contains two periods: event windows surrounding the date of profit warning and estimation period to obtain expected return by applying the model market or other models (Campbell et al. 1997).

In our case, we estimate that a period of five trading days before and after day 0 is suitable to capture the effect of profit warning while avoiding contamination from other confounding news. The date of publishing press release of profit warning is considered as the day of announcement $t_0$ or day 0.

### III-2-2 ESTIMATION OF NORMAL AND ABNORMAL RETURNS

The second step of the event study is to calculate normal and abnormal returns. The normal return is the expected gain if the event did not occur while abnormal return represents the realized return minus the normal return on the same day. In this paper, we use the market model to estimate the normal gain which assumes a linear relation between the stock return and the market return. For each security in our sample, we calculate the abnormal returns $AR_{it}$ as : $\mathbf{AR_{it} = R_{it} - E(R_{it})}$

    Where $AR_{it}$ = Abnormal returns of stock i on trading day t;
        $R_{it}$ = Actual return of stock i on trading day t;
        $E(R_{it})$ = normal return of stock i on trading day t.
    The expected return of each stock is obtained using the following formula:
        $\mathbf{E(R_{it}) = \alpha + \beta R_{mt}}$

Where α & β = are the market model parameters.

The market model parameters are estimated by ordinary least- squares regression over a 210 day period prior to the event window, from day -215 to day -5 relative to the event day, using sector index for each market sector.

Next, the average abnormal returns are calculated to measure the effect of profit warning at each time:

$$AAR_{it} = \frac{1}{N} \sum_{i=1}^{n} AR_{it}$$

Where n is the number of firms on day t.

After that, daily abnormal returns are cumulated (CAR) for each firm to measure abnormal gain for periods longer than one day:

$$CAR_i = \sum_{t=t_1}^{t_2} AR_{it}$$

Finally, CARs for the total sample are aggregated and averaged to obtain Cumulative Average Abnormal Returns:

$$CAAR : \frac{1}{N} \sum_{t=t_1}^{t_2} CAR_i$$

### III-2-3 TESTING THE SIGNIFICANCE OF ABNORMAL RETURNS IN THE EVENT WINDOW

After calculating abnormal returns around the date of profit warning, it's possible to use two types of tests, parametric and non-parametric tests to verify whether the deviation from the normal return is statistically different from zero or not.

#### III-2-3-1 PARAMETRIC TEST STATISTIC

The parametric test assumes that individual firm's abnormal returns are normally distributed and uncorrelated with each other.

The standard statistic to determine the significance of the data is:

$$T(AR) = \frac{\overline{AR_t}}{\sigma(AR_t)/\sqrt{N}}$$

Where $\overline{AR_t}$ and $\sigma(AR_t)$ are the average and standard deviation respectively of the abnormal returns of stock on day t

The t-statistics for CAR is as follows:

$$T(CAR_w) = \frac{\overline{CAR_w}}{\sigma(CAR_w)/\sqrt{N}}$$

Where $\overline{CAR_w}$ and $\sigma(CAR_w)$ are the average and standard deviation respectively of the CAR for a particular window w.

#### III-2-3-2 NON PARAMETRIC TEST STATISTIC

When the assumption of normality of abnormal returns is violated, parametric tests cause considerable over-rejection of the null hypothesis. However, non parametric tests are more powerful to detect population differences because they don't need this condition to be valid. In this paper, we will also use the Wilcoxon signed-rank test. This test is based on the idea that both sign and the extent of abnormal returns are important (Anupam 2014). Indeed, each abnormal return will be transformed into its absolute value, and then these values are ranked in ascending order (For negative variation, the absolute value will be considered). Finally, the sum of positive (negative) rank of the absolute value is calculated. The statistic is defined as:

$$\frac{T^+ - \frac{n(n+1)}{4}}{\sqrt{\frac{n(n+1)(2n+4)}{24}}}$$

Where:

        **n**: the size of the sample
        $T^+$ = the sum of positive ranks
        $T^+ = \sum_{i=1}^{n} R_{i*} * d_i$
         Where:
         $R_i$ = the rank of the variation
          $d_i$ =1 if the variation is positive
          $d_i$ =0 if the variation is negative

Under the null hypothesis of equal likely positive or negative abnormal returns and when n is large, the statistic of Wilcoxon follows a normal distribution. SPSS commands are used to calculate this test.

### IV- EMPIRICAL RESULTS

Table 3 describes the results for the full sample with eleven-day event window. The bigger market reaction is negative respectively on the announcement day (1,54%) which is statistically significant at the 1% level and the day following the profit warning (-1,17%) which is statistically significant at 5% level. This result is in alignment with previous researches in the topic that found negative abnormal return to profit warning on t=0 (Church and Donker, 2010).

To examine the difference between qualitative and quantitative warning we split our sample into two subsamples as shown in table 4.
The negative market reaction is common for both type of announcement. However, the impact is greater for qualitative warning (-1,77%) and statistically significant at 1% level, which confirms our second hypothesis.

Also, the CAAR for day 0 to day 5 (-3,492%) is huge than CAAR from day (-5) to day 0 (-1,85%), but the effect of qualitative warning around the announcement [-1, 1] is more substantial (-3,44) in comparison with the other studied event windows.

### V-CONCLUSION

This article provides evidence of market reaction around profit warning announcement of firms listed in the Casablanca Stock Exchange between the years 2009-2016. We examined whether the effect of qualitative warning is more negative compared to quantitative warnings in a short event window.
Our empirical findings show average abnormal returns on the date of announcement is negative (-1,77%) and statistically significant at the 1% level. For the total sample, the CAAR based on 5 days window surrounding the profit warning indicates significant return (-3,49%) from day 0 to day 5 at the 1% level but the CAAR prior the announcement from day -5 to day 0 is not significant. The market was not able to anticipate the profit decline; all information is integrated in the share price around the profit warning [-1, 1]. These results are consistent with findings of prior studies (Jackson and Madura 2003; Gannon 2007).
The magnitude of this negative abnormal return (-3,44%) is greater for qualitative warnings than quantitative warnings particularly in a short interval around the event because investors consider profit warnings without new numerical forecast as imprecise signal.
Finally, the effect of profit warning on share price seems to be different for each stock market. In the Casablanca Stock Market, the results provide evidence of negative market reaction for qualitative warnings but only around the announcement date.

**TABLE 1**: NUMBER OF PROFIT WARNING BY YEAR

| Years | All | Qualitative warnings | Quantitative warnings |
|---|---|---|---|
| 2009 | 1 | 0 | 1 |
| 2010 | 1 | 1 | 0 |
| 2011 | 2 | 2 | 0 |
| 2012 | 9 | 5 | 3 |
| 2013 | 3 | 2 | 1 |
| 2014 | 2 | 2 | 0 |
| 2015 | 11 | 6 | 6 |
| 2016 | 13 | 6 | 7 |
| **Total** | **42** | **24** | **18** |

*Source: press releases included in the website of the AMMC*

**TABLE 2:** NUMBER OF PROFIT WARNING BY SECTOR

| Sector | All | Qualitative warnings | Quantitative warnings |
|---|---|---|---|
| Investment companies & other finance | 4 | 4 | 0 |
| Real estate | 2 | 1 | 1 |
| Food producers | 1 | 1 | 0 |
| & processors | | | |
| Engineering & Equipment industrial goods | 2 | 1 | 1 |
| Mining | 5 | 1 | 4 |
| Leisures & Hotels | 2 | 1 | 1 |
| Construction & building materials | 4 | 3 | 1 |
| Oil & gas | 1 | 1 | 0 |
| banks | 1 | 0 | 1 |
| Holding companies | 3 | 1 | 2 |
| Materials, software &computers services | 6 | 3 | 3 |
| transport | 1 | 1 | 0 |
| Distributors | 4 | 1 | 3 |
| Chemicals | 4 | 4 | 0 |
| Forestry & paper | 2 | 1 | 1 |
| **Total** | **42** | **24** | **18** |

*Source: Casablanca Stock Exchange*

**TABLE: 3 ABNORMAL RETURNS AROUND PROFIT WARNING FOR FULL SAMPLE**

| DAYS | ABNORMAL RETURN | T-STATISTIC | WILCOXON SIGNED-RANK TEST |
|---|---|---|---|
| -5 | -0,287 | 2,461** | 0,609 |
| -4 | 0,166 | 1,882** | 0,617 |
| -3 | 0,360 | 1,854** | 0,291 |
| -2 | -0,09 | 1,850** | 0,609 |
| -1 | -0,476 | 4,749*** | 0,179 |
| 0 | -1,54 | 3,927*** | 0,002*** |
| 1 | -1,171 | 3,254** | 0,035** |
| 2 | -0,296 | 3,089** | 0,620 |
| 3 | -0,039 | 4,546*** | 0,464 |
| 4 | -0,928 | 2,412** | 0,025** |
| 5 | 0,499 | 0,809 | 0,641 |

*\*\*\* Statistical significance at the 1% level, \*\* Statistical significance at the 5% level, \* Statistical significance at the 10% level*

**TABLE 4: ABNORMAL RETURNS FOR QUALITATIVE AND QUANTITATIVE PROFIT WARNING**

| DAYS | QUALITATIVE | | | QUANTITATIVE | | |
|---|---|---|---|---|---|---|
| | AR | T-STATISTIC | WILCOXON SIGNED-RANK TEST | AR | T-STATISTIC | WILCOXON SIGNED-RANK TEST |
| **-5** | 0,249 | 2,162** | 0,626 | -1,0178 | 1,375* | 0,210 |
| **-4** | -0,058 | 1,411* | 0,732 | 0,4661 | 1,2791 | 0,286 |
| **-3** | 0,580 | 2,936** | 0,346 | 0,4011 | 2,2121** | 0,472 |
| **-2** | -0,09 | 4,067*** | 1 | -0,0855 | 0,3856 | 0,486 |
| **-1** | -0,821 | 3,043** | 0,038** | 0,015 | 3,7953*** | 0,408 |
| **0** | -1,772 | 1,612* | 0,006*** | -1,2344 | 3,673*** | 0,088* |
| **1** | -1,079 | 4,360*** | 0,046** | -1,2861 | 0,9530 | 0,316 |
| **2** | -0,931 | 1,734** | 0,224 | 0,5555 | 3,0789** | 0,500 |
| **3** | 0,435 | 2,822** | 0,186 | -0,6272 | 2,8207** | 0,744 |
| **4** | -0,823 | 1,890** | 0,064* | -1,0605 | 1,5477* | 0,199 |
| **5** | 0,146 | 2,076** | 0,773 | 0,9666 | 1,3237* | 0,286 |

*\*\*\* Statistical significance at the 1% level, \*\* Statistical significance at the 5% level, \* Statistical significance at the 10% level*

**TABLE 5:** CUMULATIVE AVERAGE ABNORMAL RETURNS DURING PROFIT WARNING EVENT WINDOWS

| DAYS | CAAR | | |
|---|---|---|---|
| | ALL | QUALITATIVE | QUANTITATIVE |
| **-5** | -0,287 | 0,249 | -1,0178 |
| **-4** | -0,121 | 0,191 | -0,5517 |
| **-3** | 0,239 | 0,771 | -0,1506 |
| **-2** | 0,149 | 0,681 | -0,2361 |
| **-1** | -0,327 | -0,14 | -0,2211 |
| **0** | -1,867 | -1,912 | -1,4555 |
| **1** | -3,038 | -2,991 | -2,7416 |
| **2** | -3,334 | -3,922 | -2,1861 |
| **3** | -3,373 | -3,487 | -2,8133 |
| **4** | -4,301 | -4,31 | -3,8738 |
| **5** | -3,802 | -4,164 | -2,9072 |

**TABLE 6:** CUMULATIVE AVERAGE ABNORMAL RETURNS OVER DIFFERENT EVENT WINDOWS

| Windows | ALL (42) | | Qualitative (24) | | Quantitative (18) | |
|---|---|---|---|---|---|---|
| **[-1,+1]** | CAAR | -3,181 | CAAR | -3,440 | CAAR | -2,835 |
| | T-STATISTIC | 4,412*** | T-STATISTIC | 2,287** | T-STATISTIC | 3,9828*** |
| | WILCOXON SIGNED-RANK TEST | 0,000*** | WILCOXON SIGNED-RANK TEST | 0,004*** | WILCOXON SIGNED-RANK TEST | 0,007*** |
| **[0,+5]** | CAAR | -3,492 | CAAR | -3,281 | CAAR | -3,773 |
| | T-STATISTIC | 3,049** | T-STATISTIC | 3,067** | T-STATISTIC | 1,525* |
| | WILCOXON SIGNED-RANK TEST | 0,006*** | WILCOXON SIGNED-RANK TEST | 0,041** | WILCOXON SIGNED-RANK TEST | 0,058* |
| **[-5,0]** | CAAR | -1,8464 | CAAR | -2,212 | CAAR | -1,359 |
| | T-STATISTIC | 1,74** | T-STATISTIC | 0,412 | T-STATISTIC | 2,911** |
| | WILCOXON SIGNED-RANK TEST | 0,16 | WILCOXON SIGNED-RANK TEST | 0,278 | WILCOXON SIGNED-RANK TEST | 0,372 |

*\*\*\* Statistical significance at the 1% level, \*\* Statistical significance at the 5% level, \* Statistical significance at the 10% level*